\documentclass[letterpaper,twocolumn,english,aps,prb,preprintnumbers,showpacs,amssymb,floatfix,showpacs,superscriptaddress]{revtex4}
\usepackage[T1]{fontenc}
\usepackage[latin1]{inputenc}
\usepackage{amsmath}
\usepackage{graphicx}
\usepackage{amssymb}

\makeatletter
\usepackage{amscd}
\usepackage{bm}

\usepackage{babel}
\makeatother
\begin{document}

\title{Non-uniform phases in metals with local moments}

\author{Angsula Ghosh}

\email{tpag2@mahendra.iacs.res.in}

\affiliation{Instituto de F\'{\i}sica Gleb Wataghin, Unicamp, Caixa Postal 6165,
Campinas SP 13083-970, Brazil}

\affiliation{Department of Theoretical Physics, Indian Association for the Cultivation
of Science, Kolkata 700032, India}

\author{E. Miranda}

\email{emiranda@ifi.unicamp.br}

\affiliation{Instituto de F\'{\i}sica Gleb Wataghin, Unicamp, Caixa Postal 6165,
Campinas SP 13083-970, Brazil}

\date{\today{}}

\begin{abstract}
The two-dimensional Kondo lattice model with both nearest and next-nearest
neighbor exchange interactions is studied within a mean-field approach
and its phase diagram is determined. In particular, we allow for lattice
translation symmetry breaking. We observe that the usual uniform inter-site
order parameter is never realized, being unstable towards other more
complex types of order. When the nearest neighbor exchange $J_{1}$
is ferromagnetic the flux phase is always the most stable state, irrespective
of the value of the next-nearest-neighbor interaction $J_{2}$. For
antiferromagnetic $J_{1}$, however, either a columnar or a flux
phase is realized, depending on conduction electron filling and the
value of $J_{2}$. 
\end{abstract}


\maketitle
The nature of the various magnetic phases of heavy fermion compounds
has been the focus of attention over the years. Most of the analysis
is based on the celebrated paradigm of Doniach, who conjectured a
phase diagram consisting of two possible phases, one paramagnetic
and another exhibiting long range antiferromagnetic order.\cite{Doniach}
The driving mechanism behind this phase diagram is the competition
between the Kondo effect,\cite{hewson} which favors paramagnetism
and is dominant at strong exchange coupling, and the Ruderman-Kittel-Kasuya-Yosida
(RKKY) interaction,\cite{rudermankittel,kasuya,yosida} which dominates
at weak coupling and can lead to antiferromagnetism. Particularly
interesting is the quantum phase transition which separates the two
phases at zero temperature and which can be accessed by tuning the
exchange interaction between local moments and conduction electrons
through external or chemical pressure. This quantum critical behavior
has been intensively studied experimentally\cite{stewartNFL,lohneysen,schroedernature}
but a complete theoretical description is still lacking.\cite{moryia,hertz,millis,mucioreview,pierspepinsirevaz,sietal}

Despite the appealing simplicity of the Doniach phase diagram the
possibility of the existence of other kinds of phases remains. Among
these we should mention inhomogeneous magnetic order,\cite{matsudaetal}
orbital antiferromagnetism\cite{chandraetal} and dimerization.\cite{dimer}
The last possibility has been given strong numerical support in the
one-dimensional case at quarter conduction electron filling.\cite{dimer}
It was ultimately ascribed to the long-ranged RKKY interaction between
localized spins.\cite{dimer} Although dimerization is an oft-encountered
instability in one dimension, its presence in higher dimensions is
less frequent. There is some (controversial) evidence in favor of
its existence in the frustrated two-dimensional Heisenberg model with
both nearest- and next-nearest-neighbor interactions.\cite{readsachdev,kotovetal,capriottisorella,capriottietal}
However, the long-ranged nature of the RKKY interaction makes its
appearance more likely in metallic systems with local moments. Motivated
by this, the aim of the present study is to look for dimerization
in particular and other forms of order with broken lattice translational
symmetry in general in higher dimensional models of heavy fermion
materials.

The co-existence of magnetic inter-site correlations and the Kondo
effect has been investigated before using mean field calculations.
Usually, two order parameters are considered: one describing the local
correlations generated by the Kondo effect and the other connected
to non-local inter-site correlations. If the inter-site correlations
break spin SU(2) symmetry, there is a competition between Kondo
singlet formation and magnetic ordering of some type.\cite{irkhinkatsjpcm,irkhinkatszpb,kimkimhong}
Alternatively, the tendency for Kondo compensation can be analyzed
in a scaling approach.\cite{irkhinkatsprb1,irkhinkatsprb2,irkhinkatsprb3}
On the other hand, if the inter-site correlations do not break spin
SU(2) symmetry, there may be the formation of some kind of spin
liquid state.\cite{colemanandrei,iglesias,ruppenthal} Fluctuations
beyond mean field have also been considered in connection with the
quantum critical behavior of the system.\cite{senthiletal,senthiletal2}
In this study, we have allowed for the emergence of broken lattice
translational symmetry in the non-local correlations, without a broken
SU(2) symmetry. We have studied the effects of conduction electron
filling and both nearest- and next-nearest-neighbor exchange interactions
on the possible phases of the Kondo lattice model in two dimensions.
The inclusion of further-neighbor interactions is intended to partially
incorporate the long-ranged nature of the RKKY interaction between
localized moments. We have found that the usually assumed uniform
state is \emph{unstable} throughout the phase diagram towards either
columnar or flux phase order.\cite{affleckmarston1,affleckmarston2}
We have also studied the temperature dependence of the order parameters.
They do not seem to differ much from the uniform case.\cite{ruppenthal}

The Kondo lattice Hamiltonian is given by\begin{equation}
H_{K}=\sum_{\mathbf{k}\sigma}\left(\epsilon_{\mathbf{k}}-\mu\right)c_{\mathbf{k}\sigma}^{\dagger}c_{\mathbf{k}\sigma}^{\phantom{\dagger}}+J_{K}\sum_{j,\alpha\beta}\mathbf{S}_{j}\cdot c_{j\alpha}^{\dagger}\bm{\sigma}_{\alpha\beta}c_{j\beta}^{\phantom{\dagger}},\label{eq:kondoham}\end{equation}
 where $\epsilon_{\mathbf{k}}$ is the band dispersion, $\mu$ is
the chemical potential, $c_{j\sigma}^{\phantom{\dagger}}$ and $c_{\mathbf{k}\sigma}^{\phantom{\dagger}}$
are conduction electron annihilation operators in real (Wannier) and
reciprocal spaces respectively, $\mathbf{S}_{j}$ is a localized spin-$\frac{1}{2}$
operator, and $\bm{\sigma}_{\alpha\beta}$ are Pauli matrices. In
addition to the above terms we also include Heisenberg-like interactions
between nearest neighbor and next-nearest neighbor localized spins
in an attempt to partially capture the long-ranged nature of the RKKY
interaction. Hence, the full Hamiltonian can now be written as $H=H_{K}+H_{H}$,
where\begin{equation}
H_{H}=J_{1}\sum_{\left\langle jk\right\rangle }\mathbf{S}_{j}\cdot\mathbf{S}_{k}+J_{2}\sum_{\left\langle \left\langle lm\right\rangle \right\rangle }\mathbf{S}_{l}\cdot\mathbf{S}_{m},\label{eq:heisenham}\end{equation}
 where $\left\langle jk\right\rangle $ and $\left\langle \left\langle lm\right\rangle \right\rangle $
denote nearest-neighbor and next-nearest neighbor sites, respectively.
In this work, we consider both ferromagnetic and antiferromagnetic
exchange interactions. The spin operators can be expressed in the
usual Abrikosov pseudo-fermionic representation\[
\mathbf{S}_{j}=\frac{1}{2}f_{j\alpha}^{\dagger}\bm{\sigma}_{\alpha\beta}f_{j\beta}^{\phantom{\dagger}},\]
 where a constraint of single $f$-electron occupancy is implied.
The mean-field Hamiltonian can be written by expressing the spin fields
in terms of the above $f$-fermionic operators and defining the following
three order parameters\begin{eqnarray}
\phi_{j\sigma} & \equiv & \frac{1}{2}\left\langle c_{j\sigma}^{\dagger}f_{j\sigma}^{\phantom{\dagger}}+f_{j\sigma}^{\dagger}c_{j\sigma}^{\phantom{\dagger}}\right\rangle ,\\
\chi_{jk\sigma} & \equiv & \frac{1}{2}\left\langle f_{j\sigma}^{\dagger}f_{k\sigma}^{\phantom{\dagger}}+f_{k\sigma}^{\dagger}f_{j\sigma}^{\phantom{\dagger}}\right\rangle ,\\
\chi_{lm\sigma}^{\prime} & \equiv & \frac{1}{2}\left\langle f_{l\sigma}^{\dagger}f_{m\sigma}^{\phantom{\dagger}}+f_{m\sigma}^{\dagger}f_{l\sigma}^{\phantom{\dagger}}\right\rangle ,\end{eqnarray}
 where $j$ and $k$ are nearest neighbor sites and $l$ and $m$
denote next-nearest neighbors. We will focus on SU(2)
invariant states, hence none of the order parameters will depend on
$\sigma$ ($\phi_{j\sigma}=\phi_{j}$, $\chi_{jk\sigma}=\chi_{jk}$,
$\chi_{lm\sigma}^{\prime}=\chi_{lm}^{\prime}$). We can write down
the mean field Hamiltonian as\begin{eqnarray}
H_{MF} & = & \sum_{\mathbf{k},\sigma}(\epsilon_{\mathbf{k}}-\mu)c_{\mathbf{k}\sigma}^{\dagger}c_{\mathbf{k}\sigma}^{\phantom{\dagger}}+E_{0}\sum_{j\sigma}f_{j\sigma}^{\dagger}f_{j\sigma}^{\phantom{\dagger}}\nonumber \\
 & - & 2J_{K}\sum_{j,\sigma}\phi_{j}\left(c_{j,\sigma}^{\dagger}f_{j,\sigma}^{\phantom{\dagger}}+\mathrm{H.\, c.}\right)\nonumber \\
 & - & J_{1}\sum_{\langle jk\rangle,\sigma}\left(\chi_{jk}f_{j,\sigma}^{\dagger}f_{k,\sigma}^{\phantom{\dagger}}+\mathrm{H.\, c.}\right)\nonumber \\
 & - & J_{2}\sum_{\langle\langle lm\rangle\rangle,\sigma}\left(\chi_{lm}^{\prime}f_{l,\sigma}^{\dagger}f_{m,\sigma}^{\phantom{\dagger}}+\mathrm{H.\, c.}\right)\nonumber \\
 & + & 4J_{K}\sum_{j}\left|\phi_{j}\right|^{2}+2J_{1}\sum_{\langle jk\rangle}\left|\chi_{jk}\right|^{2}+2J_{2}\sum_{\langle\langle lm\rangle\rangle}\left|\chi_{lm}^{\prime}\right|^{2}.\label{eq:mfham}\end{eqnarray}
 We will focus on a two-dimensional tight-binding dispersion relation
for the conduction band\begin{equation}
\epsilon_{\mathbf{k}}=-\frac{D}{2}(\cos k_{x}a+\cos k_{y}a),\label{eq:disp}\end{equation}
 where $D$ is the half bandwidth and $a$ is the lattice parameter.
The chemical potential is determined by the conduction electron density
$n$ through $\frac{1}{N}\sum_{\mathbf{k\sigma}}\langle c_{\mathbf{k}\sigma}^{\dagger}c_{\mathbf{k}\sigma}^{\phantom{\dagger}}\rangle=n$
($N$ is the number of lattice sites) and $E_{0}$ is a Lagrange multiplier
used to impose the $f$-electron single occupancy constraint on the
average $\frac{1}{N}\sum_{\mathbf{k\sigma}}\langle f_{\mathbf{k}\sigma}^{\dagger}f_{\mathbf{k}\sigma}^{\phantom{\dagger}}\rangle=1$.
The free energy can be written as\begin{eqnarray}
F & = & -2T\sum_{{\textbf{k}},\alpha=\pm}\ln[1+e^{-E_{\mathbf{k}}^{\alpha}/T}]+(E_{0}-\mu n)N\nonumber \\
 & + & 4J_{K}\sum_{j}\left|\phi_{j}\right|^{2}+2J_{1}\sum_{\langle jk\rangle}\left|\chi_{jk}\right|^{2}+2J_{2}\sum_{\langle\langle lm\rangle\rangle}\left|\chi_{lm}^{\prime}\right|^{2},\label{eq:freeenergy}\end{eqnarray}
 where $T$ is the temperature and $E_{\mathbf{k}}^{\pm}$ are the
non-interacting bands of the mean field Hamiltonian (\ref{eq:mfham}).

\begin{figure}
\begin{center}\includegraphics[%
  scale=0.4]{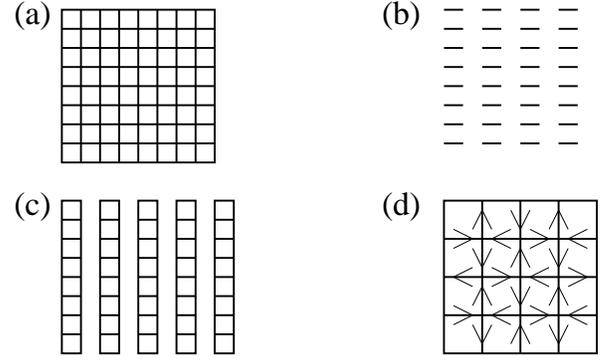}\end{center}

\caption{\label{cap:fig1}Schematic picture illustrating the various possible
phases: (a) Uniform phase, (b) Dimer phase, (c) Columnar phase, and
(d) Flux phase.}
\end{figure}

In this work, we have chosen energy and length units such that both
$D$ and $a$ are equal to 1. When translational invariance is not
broken, $\phi_{j}=\phi$, $\chi_{jk}=\chi$ and $\chi_{lm}^{\prime}=\chi^{\prime}$,
which we will henceforth call the uniform state. In addition to the
uniform case, we also consider the dimerized state with dimers along
the x-axis, the columnar phase and the flux phase.\cite{affleckmarston1,affleckmarston2}
In all cases, $\phi_{j}$ and $\chi_{lm}^{\prime}$ are taken to be
uniform. The four phases are described as follows (see Fig.~\ref{cap:fig1}):\\
 (a) Uniform: All $\chi$'s are real and equal. Lattice translation
symmetry is not broken.\\
 (b) Dimers: $\chi$ is zero for bonds along the $y$-direction whereas
for bonds along the $x$-direction we have\[
\chi_{jk}=\frac{\chi}{2}\left[1+\left(-1\right)^{j}\right].\]
 This phase has broken lattice translational and rotational symmetries.
Another state with the $x$ and $y$ directions interchanged is degenerate
and equivalent to this one.\\
 (c) Columnar: $\chi$ is uniform and equal to $\chi$ for bonds along
the $y$-direction whereas for bonds along the $x$-direction we have\[
\chi_{jk}=\frac{\chi}{2}\left[1+\left(-1\right)^{j}\right].\]
 This phase also has broken lattice translational and rotational symmetries.
By interchanging $x$ and $y$ directions we once again get another
degenerate and equivalent state.\\
 (d) Flux Phase: All of $\chi$'s are equal in magnitude but may have
imaginary phases. The specific choice of these phases is \emph{not}
gauge-invariant.\cite{affleckmarston2} However, the flux through
a plaquette is a gauge-invariant quantity. It is given by the phase
of the oriented plaquette product $\prod=\chi_{12}\chi_{23}\chi_{34}\chi_{41}$
.\cite{affleckmarston2} We consider the case depicted in Fig.~\ref{cap:fig1}(d),
in which $\prod=\pm$, staggered between adjacent plaquettes, corresponding
to fluxes of $\pm\pi$. This choice can be realized by the following
gauge choice\[
\chi_{jk}=\left|\chi\right|\]
 for bonds along the $y$-axis and\[
\chi_{jk}=\left(-1\right)^{j}i\left|\chi\right|\]
 for bonds along the $x$-axis. The bonds now being complex have a
definite direction which is shown in Fig.~\ref{cap:fig1}(d).

\begin{figure}
\begin{center}\includegraphics[%
  scale=0.3,
  angle=-90]{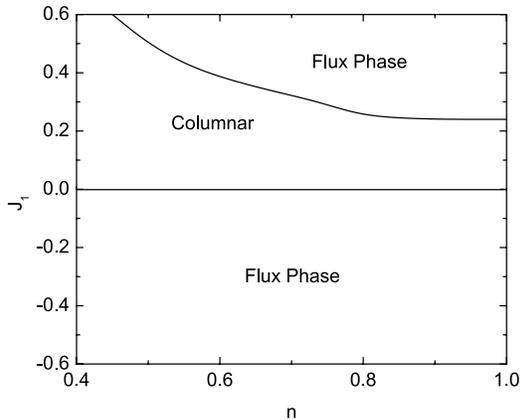}\end{center}

\caption{\label{cap:fig2}Phase diagram of the Kondo-Heisenberg model at $T=0$
as a function of the nearest-neighbor exchange $J_{1}$ and the conduction
electron filling $n$. The next-nearest-neighbor coupling $J_{2}=0$
and the Kondo coupling $J_{K}=0.5$.}
\end{figure}

We first study the phase diagram of the model at $T=0$ by varying
$J_{1}$ and $n$ while keeping $J_{2}=0$. The Kondo coupling is
kept at $J_{K}=0.5$. This is shown in Fig~\ref{cap:fig2}. We consider
both ferromagnetic and antiferromagnetic values of $J_{1}$. The first
thing to notice is the instability of the uniform state, which is
usually assumed, towards other forms of order. For antiferromagnetic
coupling between the local moments the columnar and flux phases are
the most stable, the latter occurring only for sufficiently large
$J_{1}$. However, when $J_{1}<0$ (ferromagnetic coupling), the flux
phase is the most stable, irrespective of the filling and the value
of $J_{1}$. The transition between flux and columnar phase is first
order.\\

\begin{figure}
\begin{center}\includegraphics[%
  scale=0.3,
  angle=-90]{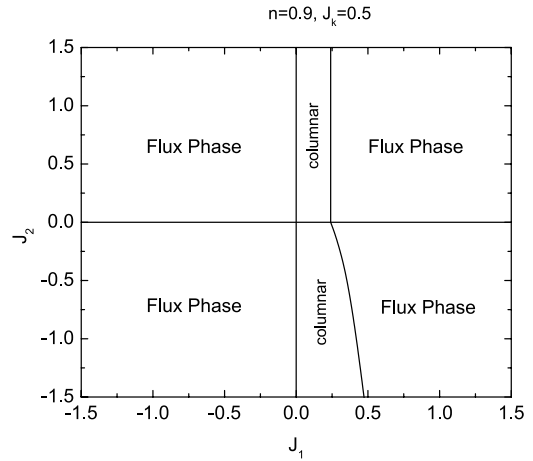}\end{center}

\caption{\label{cap:fig3}Phase diagram of the Kondo-Heisenberg model at $T=0$
with nearest-neighbor and next-nearest neighbor exchange at $n=0.9$
and $J_{K}=0.5$.}
\end{figure}

\begin{figure}
\begin{center}\includegraphics[%
  scale=0.3,
  angle=-90]{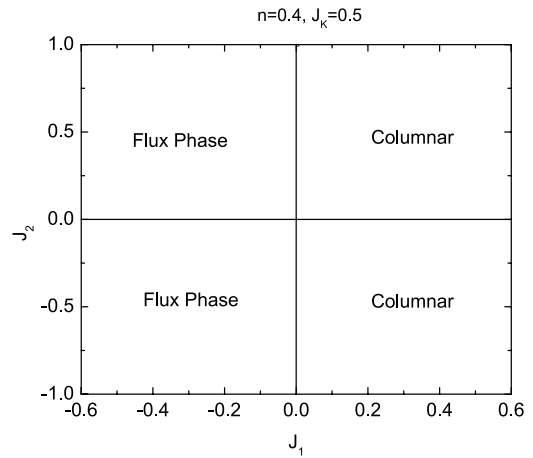}\end{center}

\caption{\label{cap:fig4}Phase diagram of the Kondo-Heisenberg model at $T=0$
with nearest-neighbor and next-nearest neighbor exchange at $n=0.4$
and $J_{K}=0.5$.}
\end{figure}

We now proceed to investigate the influence of the next-nearest-neighbor
coupling $J_{2}$ between the local moments, still at $T=0$. We studied
the phase diagram at $n=0.9$ and $n=0.4$ (Figs.~\ref{cap:fig3}
and \ref{cap:fig4}, respectively). We have allowed for both antiferromagnetic
and ferromagnetic couplings between next-nearest neighbors. For ferromagnetic
$J_{1}$, the flux phase is dominant irrespective of the value of
$J_{2}$. For antiferromagnetic $J_{1}$, columnar and flux phases
share the parameter space. For $J_{1}\leq0.24$, only the columnar
phase is realized. For higher values of $J_{1}$, a flux phase can
appear if the conduction electron filling is large enough, as shown
in Fig.~\ref{cap:fig3}. At $n=0.4$, on the other hand, the most
stable ground state is determined solely by the sign of $J_{1}$,
irrespective of the value of $J_{2}$. In this case, a ferromagnetic
$J_{1}$ favors the flux phase, whereas an antiferromagnetic $J_{1}$
leads to a columnar phase. Again, the phase boundary between flux
and columnar phases is a first order line.\\

\begin{figure}
\begin{center}\includegraphics[%
  scale=0.3,
  angle=270]{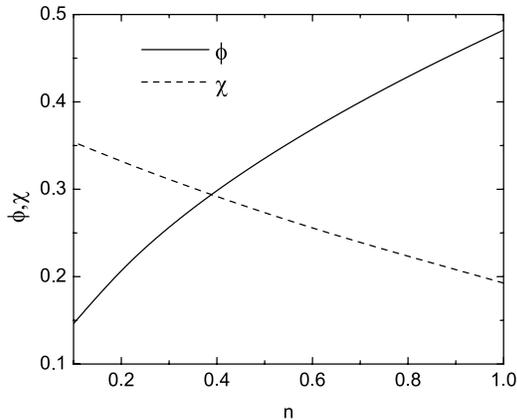}\end{center}

\caption{\label{cap:fig5}Filling dependence of the order parameters $\phi$
and $\chi$ at $T=0$ in the Kondo-Heisenberg model with nearest-neighbor
exchange only ($J_{1}=0.2$) and $J_{K}=0.5$.}
\end{figure}

In Fig.~\ref{cap:fig5}, we show the filling dependence of the order
parameters $\phi$ and $\chi$ at $T=0$, for $J_{1}=0.2$, $J_{2}=0$,
and $J_{K}=0.5$. In this case, the system is always in a columnar
phase (see Fig.~\ref{cap:fig2}). There is a clear competition between
the two types of order, the Kondo effect ($\phi$) becoming more predominant
as the system approaches half-filling. Of course, this competition
is analogous to the one predicted by Doniach between a tendency to
form to local singlets ($\phi$) and another one to lock localized
spins into some kind of order. Our mean field Ansatz is able to capture
this competition. The predominance of the Kondo effect as the system
approaches half-filling is due to an enhanced density of states in
that region providing more conduction electron states to quench the
local moments. By contrast, note that, for the same parameters of
Fig.~\ref{cap:fig5}, the uniform order parameter has a much more
reduced value and does not compete with the Kondo effect at the mean
field level (see Fig.~3 of Ref.~\cite{ruppenthal}).

\begin{figure}
\begin{center}\includegraphics[%
  scale=0.3,
  angle=270]{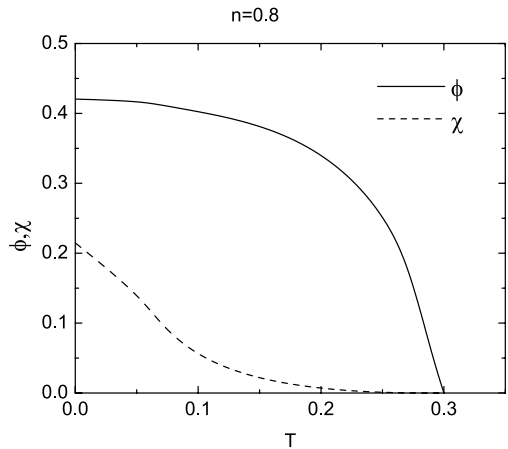}\end{center}

\caption{\label{cap:fig6}Temperature dependence of the order parameters $\phi$
and $\chi$ (columnar phase) at $n=0.8$ in the Kondo-Heisenberg model
with nearest-neighbor exchange only ($J_{1}=0.2$) and $J_{K}=0.5$.}
\end{figure}

\begin{figure}
\begin{center}\includegraphics[%
  scale=0.3,
  angle=270]{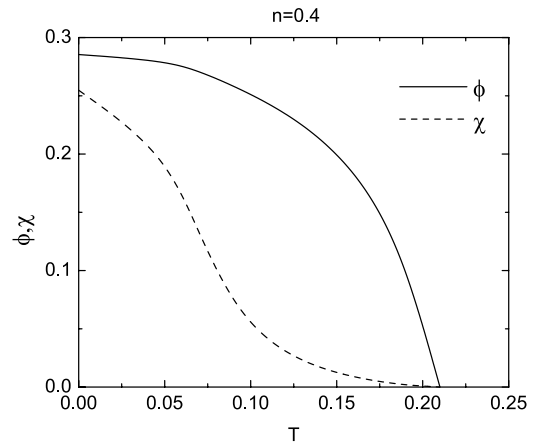}\end{center}

\caption{\label{cap:fig7}Temperature dependence of the order parameters $\phi$
and $\chi$ (columnar phase) at $n=0.4$ in the Kondo-Heisenberg model
with nearest-neighbor exchange only ($J_{1}=0.2$) and $J_{K}=0.5$.}
\end{figure}

In addition, we have also studied the temperature dependence of the
order parameters for $J_{1}=0.2$, $J_{2}=0$, and $J_{K}=0.5$. The
temperature dependence has been plotted for $n=0.8$ in Fig.~\ref{cap:fig6}
and $n=0.4$ in Fig.~\ref{cap:fig7}. In both cases, the columnar
phase is the most stable from $T=0$ up to the transition temperature.
Although the two dependences are different, both order parameters
$\phi$ and $\chi$ disappear at the same critical temperature. This
same simultaneous disappearance of order had been observed in previous
studies of the uniform phase for similar values of the exchange couplings.\cite{ruppenthal}
Although the finite temperature phase transition triggered by $\chi$
could be realized in real systems, the vanishing of $\phi$ is an
artifact of the mean field treatment.\cite{colemanlong}

Let us now pause to compare our results with previous studies. A mean
field Ansatz of the form considered here has been investigated before,\cite{colemanandrei,iglesias,ruppenthal}
without allowance for broken lattice translation symmetry. An important
conclusion of our results is that the uniform state considered in
these references is never stable. Refs.~\onlinecite{irkhinkatsprb1,irkhinkatsprb2,irkhinkatsprb3},
on the other hand, do consider the effects of both nearest and next-nearest
neighbor couplings between localized spins. Their treatment of the
Kondo effect, however, is confined to a scaling analysis, which breaks
down below the Kondo scale. Our self-consistent treatment of the Kondo
effect, by contrast, is able to reach deep into the Kondo singlet
formation regime and thus offers a better treatment below the Kondo
scale. Finally, no comparison has been attempted with the mean field
free energies of phases with conventional long-range magnetic order.\cite{irkhinkatsjpcm,irkhinkatszpb,kimkimhong}
This would determine the region of stability of these non-uniform
phases. We leave this for future studies.

In conclusion we have studied the mean field phase diagram of the
two-dimensional Kondo-Heisenberg model with both nearest- and next-nearest-neighbor
exchange interactions for various values of doping, temperature and
coupling constants. We have observed that the uniform state solution
is unstable towards lattice translational symmetry breaking for any
value of the exchange constants. Depending on the values of $J_{1}$,
$J_{2}$ and filling $n$, the system realizes either a columnar or
a flux phase. The flux phase is always stabilized by a nearest-neighbor
ferromagnetic exchange between localized spins. When this coupling
constant changes sign, however, both columnar and flux phases can
occur, the latter being favored at large $J_{1}$ and $n$ and the
former appearing at small $J_{1}$ and low fillings.

One of us (A. G.) would like to thank Dr. A. P. Vieira for helpful
discussions. The authors would like to thank the financial support of
the Brazilian Agencies FAPESP, through grants 01/00719-8 (E. M.) and
02/03799-5 (A. G.), and CNPq, though grant 301222/97-5 (E. M.) and the
Indian Agency CSIR (A. G.).

\end{document}